\documentclass{aa} 
\usepackage{times}
\usepackage{graphics}
\usepackage{epsfig}

\begin{document}

\title{A very young star forming region detected by the ISOPHOT Serendipity Survey\thanks{Based on observations with the
James-Clerk-Maxwell Telescope JCMT, the IRAM 30m Telescope, the United Kingdom Infrared Telescope UKIRT, the MPIfR 100m Telescope,
the Calar Alto Observatory
and the Infrared Space Observatory ISO, an ESA project funded by Member States (especially France, Germany, the Netherlands and the
United Kingdom) and with the participation of ISAS and NASA.} }
\author{O. Krause
\and D. Lemke
\and L. V. T\'oth
\and U. Klaas
\and M. Haas
\and M. Stickel
\and R. Vavrek
}

\offprints{O. Krause (krause@mpia-hd.mpg.de)}

\institute{
	Max--Planck--Institut f\"ur Astronomie (MPIA), K\"onigstuhl \/17, 
		69117 Heidelberg, Germany
}

\date{Received XXX; accepted XXX}
\authorrunning{O. Krause et al.} 
\titlerunning{}
\abstract{We present a multi-wavelength study of the star forming region ISOSS J 20298+3559, which
was identified by a cross-correlation of cold compact sources from the 170 $\mu$m ISOPHOT Serendipity Survey (ISOSS) database
coinciding with objects detected by the MSX, 2MASS and IRAS infrared surveys.
ISOSS J 20298+3559 is associated with a massive dark cloud complex (M $\sim$ 760 M$_{\odot}$) and located in
the Cygnus X giant molecular cloud. We derive a distance of 1800 pc on the basis of optical extinction data.
The low average dust temperature (T $\sim$ 16 K) and large mass (M $\sim$ 120 M$_{\odot}$) 
of the dense inner part of the cloud, which has not been dispersed, indicates a recent begin of star formation.
The youth of the region is supported by the early evolutionary stage of several pre- and protostellar objects discovered across the region:
I) Two candidate Class 0 objects with masses of 8 and 3.5 M$_{\odot}$,
II) a gravitationally bound, cold (T $\sim$ 12 K) and dense (n(H$_{2}$) $\sim$ 2 $\cdot$ 10$^{5}$ cm$^{-3}$) cloud core 
with a mass of 50 M$_{\odot}$ and
III) a Herbig B2 star with a mass of 6.5 M$_{\odot}$ and a bolometric luminosity of 2200 L$_{\odot}$, showing evidence for 
ongoing accretion and a stellar age of less than 40000 years.
The dereddened SED of the Herbig star is well reproduced by an accretion disc + star model.
The externally heated cold cloud core is a good candidate for a massive pre-protostellar object.
The star formation efficiency in the central cloud region is about 14 \%.
 }
\maketitle
\section{Introduction}
It is a challenge to identify massive young stellar objects during their early evolution.
The youngest protostars form deeply embedded in their cold (T $\sim$ 10-20 K) parental clouds (Pudritz 2002).
The association with dense ambient material makes such objects best detectable as cold condensations 
at far-infrared and (sub)millimeter wavelengths. The short evolutionary timescales (Palla \& Stahler 1993) 
and low spatial density of massive objects require large scale surveys for their identification.
Many of the known intermediate- and high-mass protostellar candidates have therefore been discovered by
follow-up studies towards IRAS sources (eg. Shepherd et al. 2000, Cesaroni et al. 1997, Molinari et al. 1998, Beuther et. al 2002), 
which were selected on the basis of color and flux density criteria (e.g. by Wood \& Churchwell 1989, Palla et al. 1991)

The earliest stages of massive star formation are characterized by the initial conditions of their parental cloud cores
with spectral energy distributions peaking beyond 100 $\mu$m (Evans et al. 2002).
In order to unveil such young objects we are using the ISOPHOT (Lemke et al. 1996) 170 $\mu$m Serendipity Survey (ISOSS)
(Bogun et al. 1996),
which is the largest high spatial resolution survey performed beyond the IRAS 100 $\mu$m band.
We selected bright and compact sources detected by ISOSS and IRAS with a flux ratio [F170$\mu$m/F100$\mu$m] $>$ 2, implying a low
dust temperature T $<$ 18 K and a large mass of the cold ISM in these objects.
Since the clustered mode of massive star formation commonly involves young stellar objects of different evolutionary
stages, we require the presence of embedded sources with thermal infrared excess as indicated by
the 2MASS (Cutri et al. 2000) and MSX (Price et al. 2001) infrared surveys. 
The latter criterium also avoids confusion with cold interstellar cirrus.
Here, we present the results of follow-up observations of the cold star forming region
ISOSS J 20298+3559 and show evidence for its early evolutionary stage.

\section{Observations}
\subsection{ISO far-infrared measurements}
170 $\mu$m data covering the region around ISOSS J 20298+3559 have been extracted from the ISOPHOT Serendipity Survey.
The scans performed with the C200 2x2 pixel detector array of stressed Ge:Ga with a pixel size of 89.4 arcsec 
provide a spatial resolution of 1.8 arcmin FWHM. A  flatfield correction for the individual scans was derived from 
redundant measurements at scan crossings.
The calibration accuracy is estimated to be $\pm$ 30 \%  (Stickel et al. 2000)

\begin{figure*}
\center{\psfig{figure=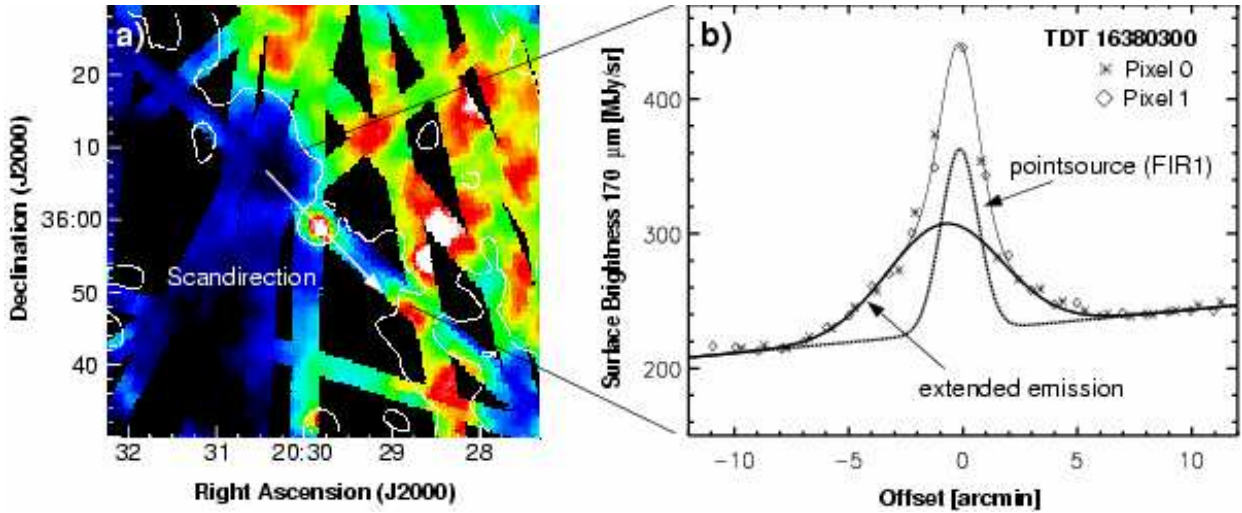,width=16.5cm,clip=true}}
\caption{
a) 170 $\mu$m continuum emission centered around ISOSS J 20298+3559,
colors are from 180 to 400 MJy/sr. Overlaid are IRAS 100 $\mu$m
contours from the HIRES processing (contour level 240 MJy/sr).
The map gives an impression of the typical sky coverage of the ISOPHOT Serendipity
Survey towards the northern galactic plane.
b) 170  $\mu$m surface brightness along
an ISOSS scan (TDT16380300) across the region (arrow in Fig. \ref{okrause1}a), reconstructed from measurements with 2 pixels of the
ISOPHOT C200 camera. The brightness profile has been decomposed into a contribution by a
compact source (FIR1) and an extended component.
}
\label{okrause1}
\end{figure*}

\subsection{SCUBA \& MAMBO (sub)millimeter mapping }
Submillimeter jiggle maps at 450\,$\mu$m and 850\,$\mu$m were obtained with the SCUBA bolometer 
array (Holland et al. 1999) at the JCMT from Mauna Kea (Hawaii) on July 28, 2001 under excellent and stable sky
conditions ($\tau_{850 \mu m}$ = 0.185 $\pm$ 0.01). The observing time was 30 min. The 
atmospheric transmission was determined from sky dips and water radiometer data (at JCMT and CSO). 
Mars, Uranus and CRL 618 served as calibrators. The data were reduced using the SCUBA User Reduction
Facility (SURF) including identification of noisy bolometer pixels and removal of sky
noise. The photometric accuracy derived from the calibration observations is 25 \% at
450 $\mu$m and 20 \% at 850 $\mu$m. The measured HPBW is 7.9 arcsec at 450\,$\mu$m and 
14.9 arcsec at 850\,$\mu$m (derived from observations of CRL618). In order to remove the error beam 
of the telescope, which contributes significantly to the 450 $\mu$m data, the SCUBA maps
were deconvolved with a beam map of Uranus using the SCLEAN algorithm (Keel 1991) and finally restored to the initial
spatial resolution.
We know that the maps may be distorted by chopping into ambient cloud emission. This will affect the 450 $\mu$m flux 
more than the 850 $\mu$m flux, because the error lobe is larger at 450 $\mu$m. The 450 $\mu$m map was therefore
finally corrected by adding a constant level of 5 \% of the peak value to the map, before convolving it to the final
resolution.

1.3mm continuum observations were carried out on June 9 and June 11, 2001 with the IRAM 30m telescope
on Pico Veleta (Spain) using the 37-channel bolometer array MAMBO (Kreysa et al. 1998). Our observing strategy adopted
On-Off measurements at 8 different raster positions in order to sample most of the object plane. The mode was
chosen due to a rather high sky opacity of $\tau_{1.2 mm}$ = 0.62 and $\tau_{1.2 mm}$ = 0.55 and unstable
atmosphere on the two days.
Calibration measurements were obtained with the nearby source K30A. Data were reduced with the NIC package including 
sky noise reduction. The photometric accuracy is 30 \%.

\subsection{UKIRT/MAX mid-infrared imaging}
Diffraction limited mid-infrared images were acquired with the MAX camera
mounted at the 3.8m United Kingdom Infrared Telescope UKIRT from Mauna Kea (Hawaii)
on December 20, 2000. The MAX camera (Robberto et al. 1998) consists of a Rockwell
128x128 Si:As BIB array, providing a field of view of 35 $\times$ 35 arcsec$^{2}$. Spectrophotometric 
data were obtained through three narrow band filters ($\lambda_{eff} \sim$  8.7, 9.7  11.6 $\mu$m,
$\Delta \lambda \sim$ 1.0  $\mu$m) and in the M-, N- and Q-bands. 
The observations were made using standard beam switching techniques with a chop frequency of 2.5 Hz and a throw of 25 arcsec 
(PA 90$^{\circ}$).
$\alpha$ Tau and $\alpha$ Boo served as standard stars, also for the narrow band filters where we
derived synthetic photometry from data obtained by Cohen (1995).
The images were finally deconvolved by the algorithm from Richardson \& Lucy with standard stars as PSF estimators and
have a resolution with a FWHM of $\sim$ 0.6 arcsec. The photometric accuracy is 20 \%.

\begin{figure*}
\center{\psfig{figure=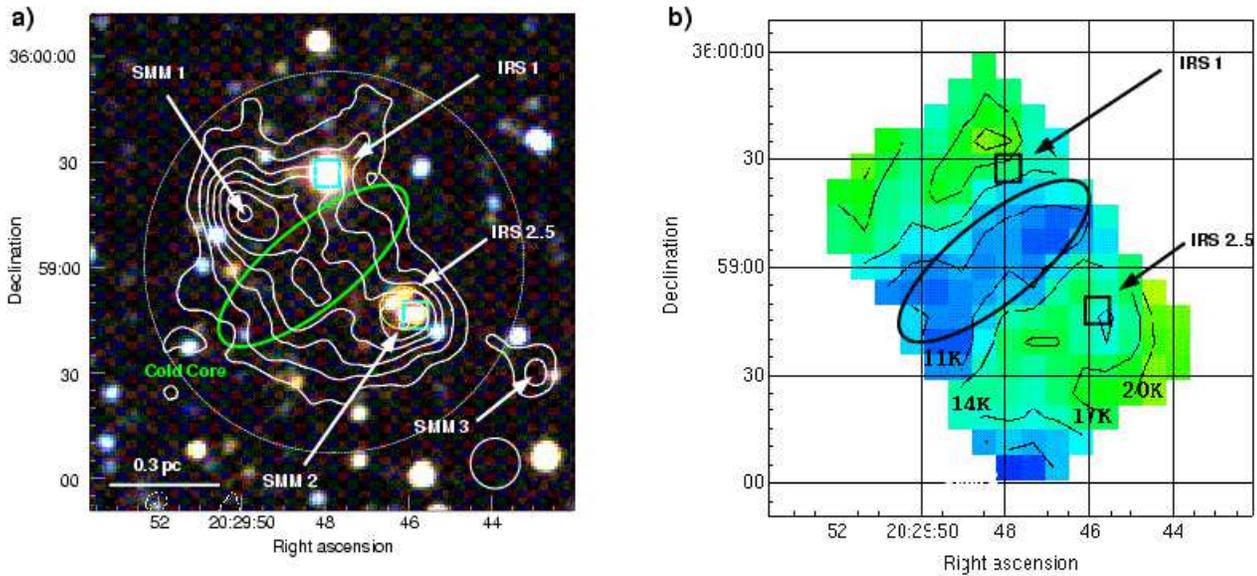,width=18cm,clip=true}}
\caption{a) 850 $\mu$m continuum map of the compact ISOPHOT Serendipity Survey source FIR1, overlaid on a near-infrared
JHK$_{S}$-composite constructed from 2MASS data.
Three compact dust condensations (SMM1, SMM2 and SMM3) are detected, which are located in a diffuse 
extended emission.
Mid-infrared sources detected by the MSX-satellite are marked with boxes.
The submillimeter knot SMM2 is associated with a small cluster of embedded NIR sources
(IRS2..5) as detected by the 2MASS and MSX surveys. IRS1 was identified as a very young Herbig B2 star 
by our follow-up spectroscopy.
The two compact submillimeter sources SMM1 \& SMM3 without any infrared counterparts are candidate
Class 0 objects. The position of the very cold cloud core is indicated by an ellipse.
Contour levels are starting at 67 and increasing by 33 mJy/beam. The size of the SCUBA beam
is indicated in the lower right, the dashed circle corresponds to the ISOPHOT beam.
b) The dust color temperature distribution across FIR1 shows the presence of a very cold (T$_{d} \sim$ 11 K) core
at the center of the cloud. The temperature profiles towards north-east and south-west indicate an external 
heating by the infrared sources IRS1-5.
The temperature is calculated from the submillimeter spectral index between 450 $\mu$m and 850 $\mu$m, 
assuming a dust emissivity $\beta$=2. }
\label{iras20278_scuba}
\end{figure*}

\subsection{Optical spectroscopy and photometry}

Optical long-slit spectroscopy was obtained using the TWIN double spectrograph at the 3.5m telescope
at Calar Alto (Spain) on December 12, 2000
with a spectral dispersion of 0.55 \AA/pixel for the blue (3400-5500 \AA) 
and 1.1 \AA/pixel in the red (6300-7100 \AA) image channel. 
A photometric accuracy of 10\,\% was obtained.  
The spectra were extracted according to the method of Horne (1986).

CCD images in the Johnson BVRI bands were obtained with the MPIA 0.7m telescope on mount K\"onigstuhl in Heidelberg
(Germany), 
equipped with a Tektronix 1024x1024 camera (scale 0.89 arcsec/pixel). 
Data reduction of both spectroscopy and direct imaging was carried out using standard
IRAF routines. I-band photometry of about 3000 stars has been derived using DAOPHOT (Stetson et. al 1987).
The stellar photometry is complete to a limiting magnitude I = 18.8 mag with
a photometric uncertainty $\sigma <$ 0.2 mag.

\subsection{Effelsberg NH$_3$ measurements}
We carried out NH$_3$ measurements with the MPIfR 100m telescope at
Effelsberg in April 2002. The HPBW at the observing frequency of
23.7\,GHz is $40\arcsec $.
The facility 1.3\,cm receiver was used with a
typical system temperature (antenna temperature units)
on the sky of 170\,K in total power mode. 
The backend was the facility
8096 channel autocorrelator split into 4 bands
in order to observe simultaneously both polarizations
at the NH$_3$(1,1) and (2,2) rest frequencies of
23694.495\,MHz and 23722.633\,MHz, respectively.
The resulting spectral resolution was 0.06\,km s$^{-1}$.
Pointing and calibration were checked by continuum
scans across NGC7027. We estimate a pointing accuracy
of about 5$\arcsec $. We calibrated our data
assuming a main beam brightness temperature for NGC7027
of 8.2 K corresponding to 5.86 Jy (Baars et al. 1977).
Our NH$_3$ spectra were reduced using
the GILDAS software, and analyzed following procedures described
by Harju et al. (1993).

\section{Results}

\subsection{Morphology of the cold ISOSS source}
Fig. \ref{okrause1} presents our far-infrared measurements of ISOSS J 20298+3559 obtained by the ISOPHOT Serendipity Survey.
The source was detected by two individual scans of the survey.
A 170 $\mu$m continuum map constructed from all scans in the region (Fig. 1a) shows the compact object on a bright
galactic background. Overlaid on the map are IRAS 100 $\mu$m contours, indicating the previous detection
of the associated source IRAS20278+3549.
Due to the superior spatial resolution of ISOPHOT we could disentangle two emission components from a
central scan across the object (Fig. \ref{okrause1}b). The 170\,$\mu$m emission is well described by an unresolved source (FIR1)
with a flux density  of S$_{170 \mu m}$=110 Jy and an extended emission component (size 5.9 arcmin FWHM,
peak surface brightness I$_{170 \mu m}$ = 90 MJy/sr).

High resolution sub-millimeter continuum maps at 450 $\mu$m and 850 $\mu$m with SCUBA and at 1.3 mm with MAMBO
have been obtained towards the unresolved far-infrared source FIR1.
The 850 $\mu$m image superimposed on a near-infrared (JHK$_{S}$) color composite from the 2 MASS survey is presented in
Fig. \ref{iras20278_scuba}.  FIR1 indicated by the dashed circle is clearly resolved by the JCMT in the submillimeter. 
The map indicates the presence of a dense
core (outlined by an ellipse) surrounded by extended emission, which covers an area of about 100 $\times$ 70 arcsec$^{2}$
and accounts for the bulk of the flux. This extended emission is located northeast and southwest of the core and shows three compact
dust condensations. We derived positions and fluxes of the compact sources by a two-dimensional gaussian
deconvolution using a local background estimation. The source sizes were determined from the 450 $\mu$m data, which have
the highest available resolution. The 1.2 mm fluxes of the fainter sources SMM2 and SMM3 could not be determined due to
background confusion at that wavelength. The results are given in Tab. \ref{submm_sources}.
The integrated (sub)millimeter flux density of FIR1 within a circular 2.4 arcmin aperture
corresponding to the SCUBA field of view and covering most of the ISOPHOT beam are
S$_{450 \mu m}$ =27.5\,Jy, S$_{850 \mu m}$ = 3.9 Jy and S$_{1200 \mu m}$ = 1.1 Jy.
\begin{table*}
\caption[]{Photometry of (sub)millimeter sources associated with ISOSS J 20298+3559-FIR1}
\begin{tabular}{l|cc|ccc|c|ccc}
\hline
\hline
Source &  Position [J2000] & & & Total Flux & & Size & & Peak Flux &  \\
     & $\alpha$ & $\delta$  &  & Jy  & & pc & & mJy beam$^{-1}$ & \\
     &          &           &  450 $\mu$m & 850 $\mu$m & 1.2 mm & & 450 $\mu$m & 850 $\mu$m & 1.2 mm \\
\noalign{\smallskip}
\hline
\noalign{\smallskip}
SMM1 & 20:29:49.9 & +35:59:15      & 1.26  & 0.233 & 0.075 & 0.1 x 0.2 & 690 & 251 & 80 \\
SMM2 & 20:29:45.6 & +35:58:46      & 0.31  & 0.091 & - & $<$0.07 & 610 & 206 & -- \\
SMM3 & 20:29:48.2 & +35:59:24      & 1.05  & 0.155 & - & 0.15 x 0.2 & 290 & 117 & - \\
cold core & 20:29:48.4 & +35:58:57 & 6.02  & 0.902  & 0.33 & 0.55 x 0.29 & 410 & 203 & 60 \\
extended  & -  & -         &18.86  & 2.47  & 0.7 & 0.9 x 0.7 & 368 & 173 & - \\
\hline
\noalign{\smallskip}
\end{tabular}
\label{submm_sources}
\end{table*}

\subsection{Distance and structure of the cloud complex}
\begin{figure}
\center{\psfig{figure=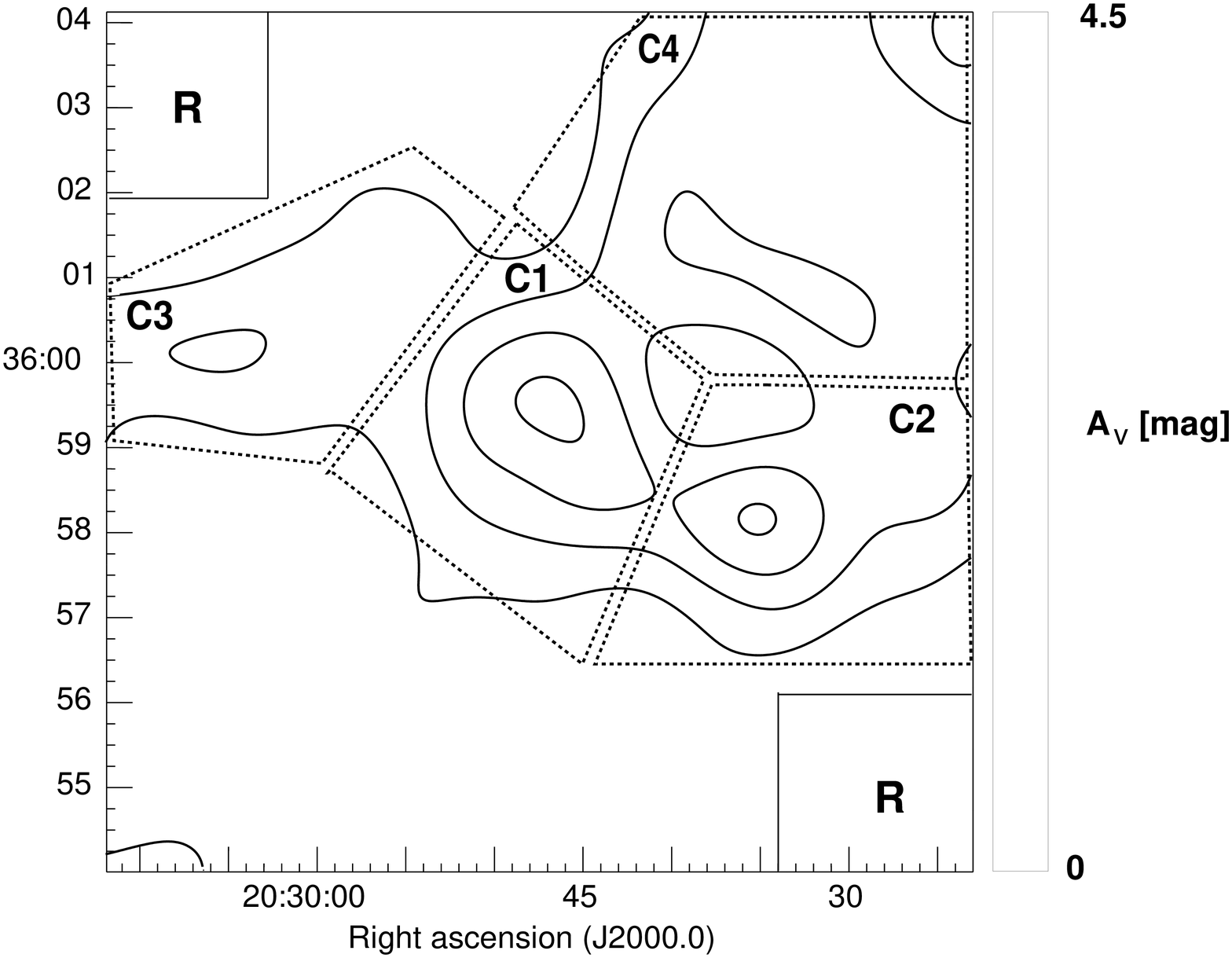,width=\linewidth,clip=true}}
\caption{
Extinction map towards the star forming region ISOSS J 20298+3559 derived from I-band star counts and
corrected for foreground stars. Four individual dark clouds (C1..C4) have been detected. Overlaid are contours
of visual extinction A$_{V}$ (contour levels at 1.5, 2.5, 3.5 and 4.5 mag). The rectangles marked with 'R'
show the reference fields used to set the zero point of extinction. In order to obtain the total
line-of-sight extinction, a foreground extinction of A$_{V}$ = 2.7 mag has to be added to the data.
}
\label{extinction_segments}
\end{figure}
A well known drawback for the use of kinematical distances towards Cygnus is the near-zero radial velocity gradient up to a distance
of about 2 kpc from the sun. Therefore we have created an extinction map to determine a reliable distance to ISOSS J 20298+3559 based 
on a Wolf diagram and to study the 
large-scale distribution of dust associated with the far-infrared source. We have obtained deep I-band images of the region using the 
MPIA 0.7m telescope.

\begin{figure}
\center{\psfig{figure=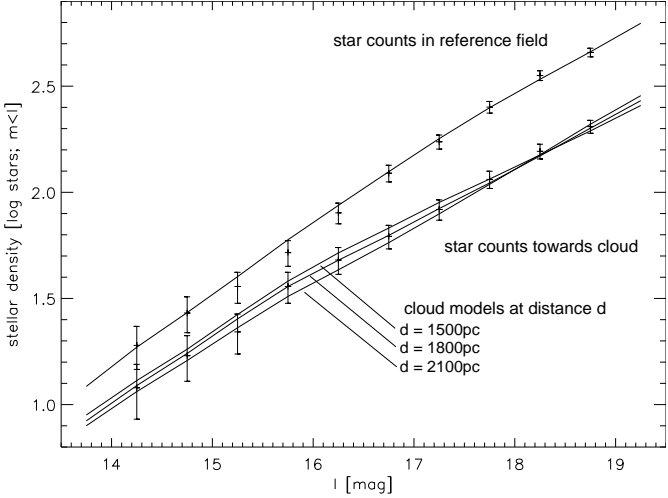,angle=0,width=\linewidth,clip=true}}
\caption{Wolf-diagram of cumulative I-band star counts towards the dark cloud (lower points) and the reference fields (upper points).
The error bars correspond to $\pm 1 \sigma$ of a Poisson statistics.
The data are in good agreement with our predictions based on a Galactic stellar distribution model, indicated by the
overlaid curves. The results of the Monte-Carlo simulations are consistent with a distance of 1.8 $\pm$ 0.3\,kpc to the dark cloud complex }
\label{extinction2}
\end{figure}

The extinction map shown in Fig. \ref{extinction_segments} was constructed according to the method of Bok (1956) and
subsequently smoothed to the ISOPHOT resolution.
An adaptive grid (reseau size $\sim$ 1 arcmin) was used for optimization of the angular resolution with respect to the
stellar density (Cambr\'esy et al. 1997). The observed cloud extinction $A^{obs}$ with respect to the reference positions
was calculated from the number density $N^{b}$ of stars seen towards the cloud and their unobscured
reference density  $N_{ref}^{b}$: $A^{obs}= -1/$a log(N$_{ref}^{b}/N^{b})$, where
$a$ denotes the slope of the I-band luminosity function.
We use A$_{V}$ = 2.07 A$_{I}$ (Rieke \& Lebofsky 1985) to relate visual extinction to our I-band observations.
The reference fields are marked with 'R' in Fig. \ref{extinction_segments} and are located in direction of the lowest dust column 
density as indicated by the 170 $\mu$m continuum.

The extinction $A^{obs}$ derived from the simple Bok formula underestimates the true cloud
extinction $A^{true}$ of a distant cloud due to the contamination by foreground stars.
Foreground stars lower the density contrast towards the cloud. Since they are not affected by the cloud
extinction, their number density remains constant: $N^{f}$ =  $N_{ref}^{f}$.
In order to account for the stellar foreground
and background population we apply the general form of the Bok formula (e.g. Cambr\'esy et al. 2002)
providing the true extinction $A_{I}^{true}$ depending on the fraction of background stars.

Fig. \ref{extinction2} presents the Wolf diagram for the reference field (upper curve) and towards the dense region 
($A_{I}^{obs} >$ 1 mag) of the cloud
complex (lower curve). 
The logarithm of the cumulative 
distribution of apparent magnitudes is shown, error bars are $\pm$ 1 $\sigma$ and are calculated according to Poisson statistics.
The number density in the reference field agrees well with the prediction from the Galactic
stellar distribution model from Robin \& Creeze et al. (1986), assuming an average interstellar extinction coefficient of $A_{V}$ = 1.5 mag kpc$^{-1}$
in the distance range 0 $<$ d $<$ 4 kpc and $A_{V}$ = 0.7 mag kpc$^{-1}$ beyond 4 kpc. The higher extinction
below 4 kpc is attributed to dust in several giant molecular cloud complexes of the local spiral arm
and extending out to the inter-arm gap between local and Perseus spiral arms at a distance of 4 kpc (Ungerechts et al. 2000).
The integrated (0 $<$ d $<$ 5 kpc) line-of-sight extinction of $A_{V}$ = 6.7\,mag towards the reference fields agrees with the
value of $A_{V}$ = 6.9 $\pm$ 0.7 mag derived by Schlegel \& Finkbeiner (1998) from the IRAS 100 $\mu$m dust continuum.

To interprete the shape of the Wolf diagram correctly, a model curve of an obscured region was determined from Monte-Carlo simulations
using a Galactic stellar distribution model for the unobscured field star luminosity function.
We assumed the extinction in the cloud to be gaussian distributed for resolutions smaller than tested by the average extinction A$_{I}^{obs}$ 
of our integral aperture.
The lower curves in Fig. \ref{extinction2} show the results of our simulations for three best-fit cloud models, which are
consistent with our observed average extinction ($A_{I}^{obs}$ = 1.1 mag) and are located at distances of 1.5, 1.8 and 2.1 kpc,
respectively.
From the observations we derive a cloud distance d = 1.8 $\pm$ 0.3 kpc and an average extinction of the cloud of $A_{V}$ = 3.1 $\pm$ 0.5 mag,
where the errors mainly originate from the Poisson statistics of our star counts. 
Using the interstellar extinction coefficient of $A_{V}$ = 1.5 mag kpc$^{-1}$ determined from the reference field, the 
foreground extinction towards ISOSS J 20298+3559 is  $A_{V}$ = 2.7 $\pm$ 0.3 mag.

We compared our result with the empirical distance estimator d = 320 N$^{0.57}$ pc by Herbst \& Saywer (1987), which relates
the cloud distance to the number N of foreground stars counted within a circular aperture of 5 arcmin diameter on the POSS blue print.
Using B-band star counts from the USNO-PMM catalogue (Monet 1996) we obtain a distance of d $\sim$ 2.2 kpc for the cloud complex.
The formula is however only valid for opaque dark clouds (A$_{V} > $ 7 mag) and the higher distance might originate
in the partly translucent character of ISOSS J 20298+3559.

\subsection{Cloud masses}
The extinction map clearly reveals the presence of a dusty cloud complex with 4 different dark clouds (C1..C4).
C1 coincides with the cold source FIR1 and accounts for the extended far-infrared emission shown in Fig. \ref{okrause1}b. 
Despite being covered by our 170 $\mu$m scan, the south-western cloud C2 does not show up with a strong 170 $\mu$m excess 
and we find an upper limit of S$_{170 \mu m}$ $<$ 15 Jy.
The cloud complex also shhows a small extinction ridge towards the east (C3) and extends to 
the north-west (C4).

Cloud masses were derived from the extinction map, using the relation by Dickman (1978):
\begin{equation}
M = \left ( \Delta \Omega d \right ) ^{2} \mu \frac{N_{H}}{A_{V}} \sum_{i} A_{V}(i),
\end{equation}
where $\Delta \Omega$ is the angular size of the map, d is the distance to the cloud, $\mu$ is the mean molecular
weight corrected for helium abundance, and i corresponds to a map pixel.
With the dust-to-gas ratio proposed by Savage \& Mathis (1979), $N_{H} / A_{V} = 1.87 \times 10^{21} $cm$^{-2} $mag$^{-1}$
($N_{H} = N_{H I} + 2 N_{H_{2}}$) we obtain a total mass of the cloud complex of 760 M$_{\odot}$. We list in Tab. \ref{clouds}
parameters of the four individual clouds. Due to systematic uncertainties in the extinction values and the conversion factor, we 
assume the mass estimates to be accurate within a factor of 2. 

\begin{table}
\caption[]{Individual clouds towards ISOSS J 20298+3559.
The columns are: (1) cloud, (2) position of peak extinction A$_{V}$, (3) mass M$_{H I}$ +
M$_{H_{2}}$ , (4) peak extinction A$_{V}$ and (5) size.
}
\begin{tabular}{l|c|c|c|c}
\hline
\hline
Source &  Position [J2000] & Mass & Peak A$_{V}$ & Size \\
     & $\alpha$, $\delta$  & M$_{\odot}$ & mag  & pc$^{2}$\\
\noalign{\smallskip}
\hline
\noalign{\smallskip}
C1 & 20:29:46.8  +35:59:30 & 188  & 4.7 & 2.0 $\times$ 1.7 \\
C2 & 20:29:35.1  +35:58:09 & 162  & 4.6 & 1.9 $\times$ 1.7 \\
C3 & 20:30:05.3  +36:00:07 & 115  & 2.6 & 2.5 $\times$ 1.5 \\
C4 & 20:29:31.9  +36:00:49 & 295  & 3.7 & 2.3 $\times$ 2.2 \\
\hline
\noalign{\smallskip}
\end{tabular}
\label{clouds}
\end{table}

\subsection{Dust properties of FIR1}
\begin{figure}
\center{\psfig{figure=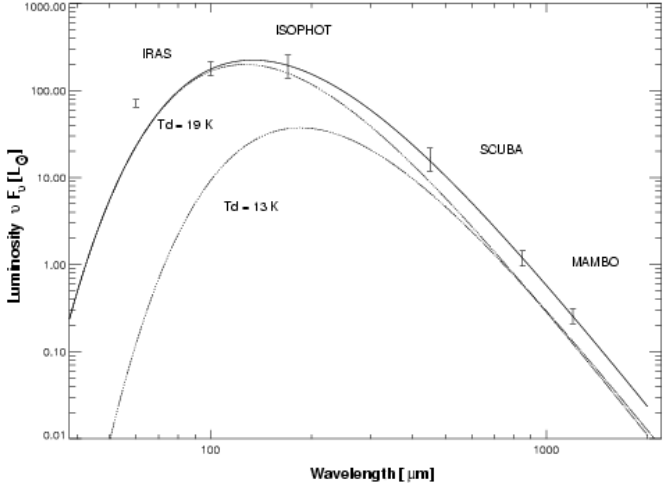,width=\linewidth,clip=true}}
\caption{Spectral energy distribution of the cold ISOSS source FIR1.
The total flux density is well characterized by the superposition of optically thin thermal radiation from two
modified blackbodies ($\beta$ = 2) with dust temperatures of T$_{d}$ = 13 $\pm$ 2 K and T$_{d}$ = 19 $\pm$2 K,
respectively.}
\label{cold_bbs1}
\end{figure}

Fig. \ref{cold_bbs1} displays the spectral energy distribution (SED) for the cold region FIR 1.
The spectral energy distribution longwards of 100 $\mu$m is dominated by large grains, which are at an
equilibrium temperature within the surrounding radiation field. 

%%%%%%%%%%%%%%%%%%%%%%
\begin{table*}
\caption{Results of the ammonia observations of the cold core region. The columns are
(1) line, (2) main beam brightness temperature of the maingroup, (3) v$_{LSR}$, (4)
FWHM from fit, (5) main group line area, (6) total line area of satellite groups
and (7) $\tau_{main}$.}
\begin{center}
\begin{tabular}{l|c|c|c|c|c|c}

\hline
\hline
Line  & $T_{\rm MB}$(main)  & $v_{\rm LSR}$ &  FWHM         &  areamain   & areasatel  & $\tau $ \\
      & K                   & kms$^{-1}$     &  kms$^{-1}$    &  Kkms$^{-1}$ & Kkms$^{-1}$ &    \\
\hline
(2,2) & 0.139 (0.074)       & 0.66(0.13)    &  0.883(0.408) &   0.146     &            & 0.78(0.10)\\
(1,1) & 0.329 (0.076)       & 0.58(0.02)    &  0.955(0.065) &   0.34      & 0.19       & 0.10(0.05)\\

\hline
\end{tabular}
\end{center}
\label{ammonia_results}
\end{table*}

The flux density of the diffuse dust component in FIR1 (excluding the knots SMM1, SMM2 and SMM3) is
 well characterized by optically thin thermal radiation of two
modified blackbodies $S_{\nu} \propto \nu^{\beta} B(\nu, T_{d})$ with a dust emissivity $\beta$=2 and dust temperatures 
T$_{d}$ = 13 $\pm$ 2 K and T$_{d}$ = 19 $\pm$2 K. 
As can be seen from Tab. \ref{submm_sources}, the contribution of SMM1, SMM2 and SMM3 to the total (sub)millimeter flux
is less than 15 \%. In the large beam measurements of the IRAS/ISO satellites we cannot
separate these sources. If they contributed significantly more FIR flux than estimated from our submillimeter data, the dust temperatures
of the diffuse dust component would be even lower and the resulting dust mass higher.

Drawing the conclusion of a very cold dust component merely from large-beam integrated photometry is critical without any radiative
transfer model. Our spatially resolved submillimeter data allow, however, to confirm the presence of
a very cold dust component from the dust color temperature map of ISOSS J 20298+3559-FIR1 presented in Fig. \ref{iras20278_scuba}b.
The map was obtained from the deconvolved 450 $\mu$m and 850 $\mu$m maps smoothed to 14.9 arcsec
resolution. The color temperature T$_{col}$ was derived from the spectral index between the two wavelengths and is given by
\begin{equation}
F_{\nu_{1}} / F_{\nu_{2}} = \nu_{1}^{3+\beta} (e^{h \nu_{2} / k T_{col}} -1 ) / \nu_{2}^{3+\beta} (e^{h \nu_{1} / k T_{col}} -1 ),
\end{equation}
with a dust emissivity $\beta$ = 2.
A decreasing color temperature and a clear 850 $\mu$m (and 1.2 mm) excess is observed towards the cloud center,
exhibiting a minimum temperature of only T$_{col}$ = 11 $\pm$2 K.

Optical depth effects in the sub-millimeter or a change in the dust emissivity ($\beta < 2$) could affect the
spectral index. We conclude, however, that these effects are small and the dust temperature dominates the spectral index:
The peak column density towards the cloud core derived from our 850\,$\mu$m data
is N(H$_{2}$) = 2.2 $\cdot$ 10$^{22}$ cm$^{-2}$, in good agreement with our ammonia observation (see next section). The
corresponding optical depth is $\tau_{850}$ = 2 $\cdot$ 10$^{-3} << 1$. Assuming a roughly ellipsoidal core geometry
of $0.55 \times 0.29$ pc$^{2}$
as indicated in Fig. \ref{iras20278_scuba} we obtain an absolute molecular gas density of  N(H$_{2})$ = 2 $\cdot$ 10$^{5}$ cm$^{-3}$. 
Concerning $\beta$, the centrally decreasing temperature profile points to an external heating of the core region.
Therefore the conditions in our core region basically resemble observations of nearby very cold pre-stellar cores 
for which $\beta$=2 was found (eg. Juvela et al. 2001, Ward-Thompson et. al. 2002). 

Masses for the emitting regions have been derived from the optically thin emission in the submillimeter. The dust mass is given
by M$_{d}$ = $F_{850} d^{2} / (\kappa_{850} B_{850}(T_{d}))$,
where we used a dust mass absorption coefficient of $\kappa_{850}$ = 2.1 cm$^{2}$ g$^{-1}$ following Ossenkopf \& Henning (1997) for
dense protostellar cores. Following Sodroski et al. (1997) we assume a gas-to-dust mass ratio of 150 
and derive a total gas mass M$_{H_{2}}$=120 M$_{\odot}$ $\pm$
24 M$_{\odot}$for FIR1. The total mass contained in the cold core is M$_{H_{2}}$ = 50 $\pm$ 10 M$_{\odot}$.

We calculated the bolometric FIR luminosities re-radiated from large grains by integrating over the Planckian radiation
and find for the region L$_{FIR}$ = 250 L$\odot$ and for the cold core L$_ {FIR}$ = 15 L$\odot$, respectively.

\subsection{Dense gas in the cold core region}

It is important to approve the physical conditions derived
from our dust continuum observations by molecular line observations, which
reflect the physical conditions of the gas phase ISM.
NH$_3$ is a very useful indicator of molecular cloud temperatures and
column densities (e.g., Walmsley \& Ungerechts 1983).

\begin{figure}
\center{\psfig{figure=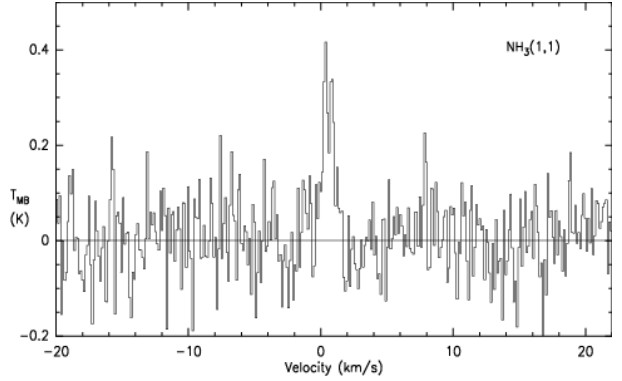,width=8cm,clip=true}}
\caption{Effelsberg $NH_{3}$ spectrum taken at the position of the cold core region (Table 1) in ISOSS J 20298+3559-FIR1}
\label{effel_nh3}
\end{figure}

\begin{figure*}
\psfig{figure=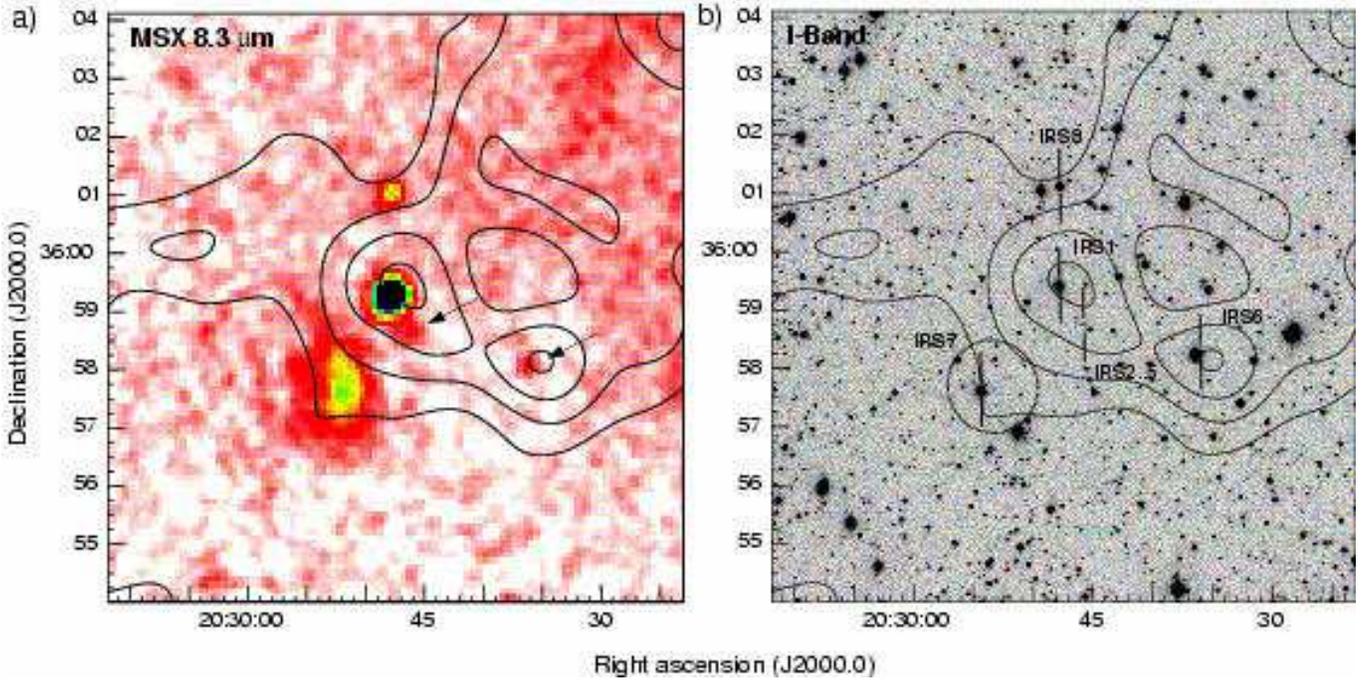,width=\linewidth,clip=true}
\hfill
\caption{
a) 8.3 $\mu$m image obtained by the MSX satellite with contours of visual extinction overlaid
(see Fig. \ref{extinction_segments}, contour levels A$_{V}$=1.5, 2.5, 3.5 and 4.5 mag). 
Colors are stretching from 5 to 40 mJy pixel$^{-1}$. Five sources are detected by MSX above
the 5 $\sigma$ level,  the two fainter ones are marked by arrows. Note the extended source towards the south-east, which shows an
external illumination of the outer rim of the molecular cloud. The illuminating star is IRS7.
b) I-Band image of the ISOSS J 20298+3559 region, obtained with the MPIA 0.7m telescope with contours of visual extinction 
overlaid. The optical identifications of mid-infrared sources seen by the MSX satellite are labeled. 
}
\label{ir_overlays}
\end{figure*}

We detected the NH$_3$(1,1) line with a S/N of 4 and the NH$_3$(2,2) line with a S/N of 2. Fig. \ref{effel_nh3}
presents the (1,1) spectrum. We derived the NH$_3$(1,1) line optical depth $\tau_{11}=0.10\pm0.05 $ and
estimated the (1,1)(2,2) rotational temperature $T_{\rm 12}=19.2\pm4.0$ K and
the gas kinetic temperature $T_{\rm kin}=21.3\pm 5.0$ K. The derived ammonia
column density was estimated as N(NH$_{3}$)=1.32$\pm 0.24\times 10^{14}$cm$^{-2}$.
From the $N({\rm H}_2)$ column density obtained by our submillimeter continuum
observations and smoothed to 40 arcsec resolution we derive a molecular hydrogen
column density of $N({\rm H}_2)= 1.3\pm 0.6\times 10^{22}$cm$^{-2}$ for the region
covered by the Effelsberg beam. This corresponds to a fractional abundance of
$\chi ( NH_{3}) = N(NH_{3}) / N(H_{2})$ = 1.0 $\cdot$ 10$^{-8}$ which is in the
range found by Molinari et al. (2000) in their study of ammonia clumps associated
with young intermediate and high mass star forming regions.
Ammonia in ISOSS J 20298+3559 seems however to be under-abundant when comparing to the
values given by Harju et al. (1993). They find a mean value of
$\chi ( NH_{3}) \sim 2.6  \cdot 10^{-8}$ in a sample of 22 ammonia clumps in Orion.
Ammonia is considered as a molecule characterizing later stages of chemical evolution.
Myers \& Benson (1993) found $NH_{3}$ to be more abundant in older cores, where stars have
already formed. This may indicate that ISOSS J 20298+3559 is a rather young object.
Tab. \ref{ammonia_results} summarizes our ammonia results.

ISOSS J 20298+3559 is located towards the southern outskirts of the Cygnus X Giant Molecular Cloud (Dame \& Thaddeus 1985).
The majority of active and luminous star forming regions in this giant molecular cloud is found at
distances between 1.2 and 2.5 kpc (Odenwald \& Schwartz 1993), corresponding to radial velocities (LSR) between -10 km s$^{-1}$ and
+10 km s$^{-1}$. The ammonia line velocity v$_{LSR}$ = 0.58 $\pm$ 0.06 km s$^{-1}$ is consistent with the
kinematical association with the Cygnus X GMC.
\begin{figure}
\hspace{-0.5cm}
\psfig{figure=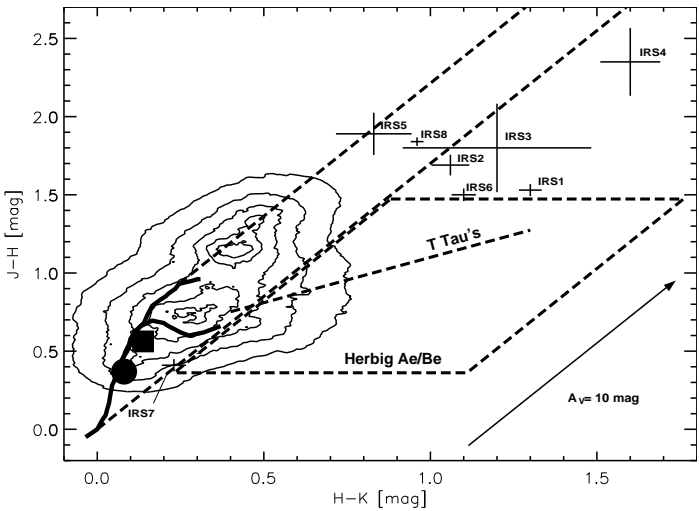,width=9.2cm,clip=true}
\caption{
2MASS JHK$_{S}$ color-color diagram towards the cloud complex. The number density of field stars is
indicated by contours. The solid curve in the lower left of the diagram is the locus of unreddened main
sequence stars and giants, their reddening band is indicated between the two dashed lines.
Several of the MSX counterparts IRS1..8 show an infrared excess characteristic for young stellar objects,
as indicated by the colors of Herbig Ae/Be
and T Tauri stars. Also given is the redenning vector from Rieke \& Lebofsky (1985).
}
\label{ir_colors}
\end{figure}
\begin{table*}
\caption[]{Infrared sources associated with the cloud complex
The columns are: (1) IR source label, (2) position (3-6) near-infrared brightness in IJHK$_{S}$ bands,
(7) 8.3 $\mu$m flux, (8) 2MASS PSC identification, (9) additional remarks.
}
\begin{tabular}{l|c|c|c|c|c|c|c|l}
\hline
\hline
Source &  Position [J2000] & I & J & H & K$_{S}$ & 8.3 $\mu$m & 2MASS & remarks\\
     & $\alpha$, $\delta$  & \multicolumn{4}{c}{mag} & mJy & counterpart & \\
\noalign{\smallskip}
\hline
\noalign{\smallskip}
IRS1 & 20:29:47.9 +35:59:26 & 12.90 & 10.98 & 9.45 & 8.15 & 1740 & J2029479+355926 & Herbig B2 star \\
IRS2 & 20:29:45.8 +35:58:47 & 17.23 & 14.70 & 13.01 & 11.95 & 90 & J2029458+355847 & brightest member of cluster IRS2..5\\
IRS3 & 20:29:46.1 +35:58:46 & 18.54 & 16.1$^{\star}$ & 14.3$^{\star}$ & 13.1$^{\star}$ & $|$ & & JHK$_{S}$ fluxes from 2MASS map\\ 
IRS4 & 20:29:46.1 +36:58:52 & 19.5  & 16.2$^{\star}$ & 13.85 & 12.25 & $|$ & J2029461+365852 & J flux from 2MASS map\\ % bright K for northern couple
IRS5 & 20:29:46.4 +35:58:50 & 18.44 & 15.74 & 13.85 & 13.02 & $|$ & J2029464+355850 &\\
IRS6 & 20:29:36.0 +35:58:18 & 15.77 & 13.33 & 11.83 & 10.73 & 75 & J2029360+355818 & \\
IRS7 & 20:29:54.5 +35:57:38 & 12.48 & 11.38 & 10.97 & 10.74 & 1870 & J2029545+355738 & MIR nebula + illuminating star\\
IRS8 & 20:29:47.9 +36:01:08 & 13.09 & 9.78 & 7.94 & 6.98 & 225 & J2029479+360108 & background star\\
\hline
\noalign{\smallskip}
\end{tabular}
\label{ir_sources}
\end{table*}

\subsection{Young stellar objects in the cloud complex}
Mid-infrared observations are well suited to identify young stellar objects
enshrouded by warm circumstellar dust.
An MSX 8.3 $\mu$m image covering the region of our extinction map is presented in Fig. \ref{ir_overlays}a.
Four compact and one extended mid-infrared sources have been detected above the 5 $\sigma$ noise level ($\sigma$ $\sim$ 15 mJy/beam).
The astrometric accuracy of the MSX image ($\sim$ 2 arcsec RMS) allows the cross-identification with objects from
our I-band image shown in Fig. \ref{ir_overlays}b and the 2MASS point source catalogue. 
All four compact MSX sources coincide in position with very red (I-K$_{S}$ = 5..7 mag) 
near-infrared counterparts. The objects are marked in Figs. \ref{iras20278_scuba} and \ref{ir_overlays}b. 
Astrometry and photometry of the sources are given in Tab. \ref{ir_sources}, where
the 8.3 $\mu$m fluxes were derived from the MSX map performing aperture photometry, yielding a photometric accuracy of 25 \%.

The intrinsic excess emission and extinction of the MSX near-infrared counterparts are characterized by
the (J-H) vs. (H-K$_{s}$) color-color diagram presented in Fig. \ref{ir_colors}, where the colors of IRS1..8 are
plotted with their photometric 1\,$\sigma$ errors. The solid curve corresponds to the loci of unreddened main sequence
and giant branch stars (Bessel \& Brett 1988) and is extended by the reddening band, which confines stars with normal
photospheres (dashed parallel lines).

IRS1, IRS2, IRS3, IRS4, and IRS6 lie to the right side and show an intrinsic infrared excess due to warm circumstellar 
dust. Their observed colors are consistent with reddened (A$_{V} \la$ 15 mag) pre-main-sequence stars of 
low- to intermediate-mass as indicated in the diagram by the loci of unreddened Herbig Ae/Be (dashed parallelogram, 
Hillenbrand et al. 1992) and T Tauri stars (dashed line, Meyer et al. 1997).
The derived extinction is smaller than the total line-of-sight extinction derived from our 850 $\mu$m, yielding
A$_{V}$=18 mag towards IRS1 and  A$_{V}$=23 mag towards the small cluster IRS2, IRS3, IRS4 and IRS5.
The sources are not affected by the total dust column, and are therefore either foreground objects or embedded in the cloud complex. The positional coincidence of 
dust with IRS1..5 (Fig. \ref{iras20278_scuba}) favors however the association with the dark cloud:
IRS1 is located on a dust ridge in the northwest of the central region of the cloud C1, IRS2 coincides in position
with the compact submillimeter source SMM2 and is surrounded by a compact cluster of a least 3 further red sources,

IRS5, IRS7 and IRS8 show no infrared excess. IRS5 and IRS8 are reddened with about (A$_{V} \sim$ 15 mag).
The corresponding line-of-sight extinctions from  our extinction maps are 23 mag for IRS5 and 4.5 mag for IRS7 and IRS8. 
which are located at the border of the cloud. This suggests that also IRS5 and IRS7 are associated with the cloud, while IRS8 is a background giant.

IRS7 is located approximately in the center of a spherically segmented mid-infrared nebula seen in Fig. \ref{ir_overlays}a. 
The nebula is located in the south-east of the dark cloud C1 and indicates an external illumination of the outer rim of the 
molecular cloud by IRS7. The nebula and IRS7 coincide in position with the source IRAS20279+3547. 
The flux ratio F12/F25 $>$ 1.4 and the strong emission in the MSX 8.3 $\mu$m band, which contains a strong PAH feature,
indicate a reflection nebula around a 
B type star. In order to clarify whether IRS7 is the illuminating source we have obtained
BVRI photometry of the source. Comparing with the colors of main sequence stars we obtain a best fit
for A$_{V}$ =  4.2 $\pm$ 0.3 mag and a spectral type B4..B6, in agreement with the total line-of-sight extinction  A$_{V}$ =  4.2 $\pm$ 
0.5 mag towards IRS7 from our extinction map.

\begin{figure*}
\psfig{figure=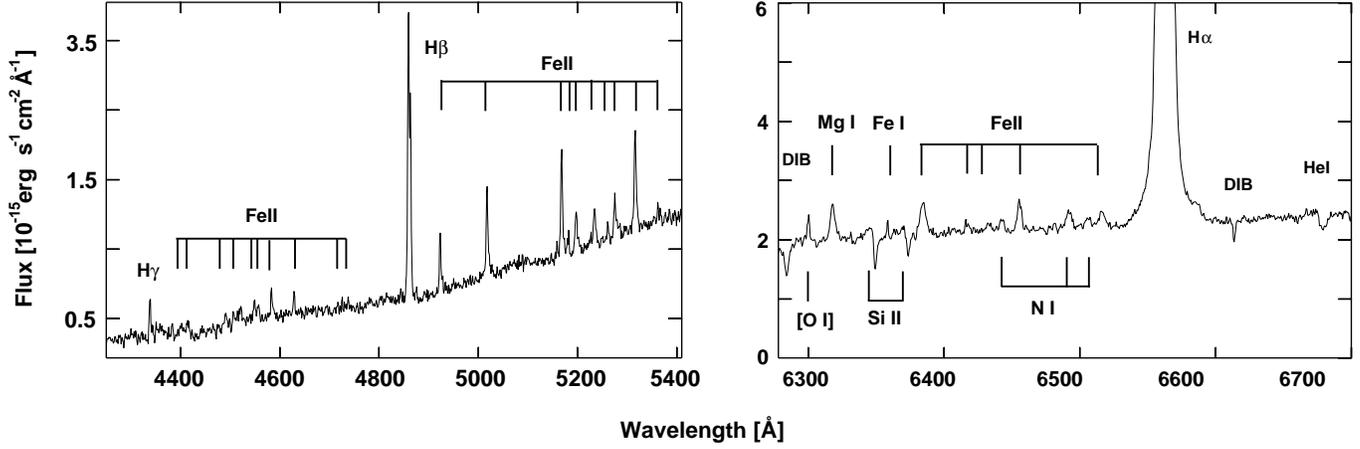,width=\linewidth,clip=true}
\caption{Optical spectrum of the YSO ISOSS J 20298+3559-IRS1 taken with the TWIN spectrograph at the 3.5m
telescope at Calar Alto. A spectral type B2 was inferred from photospheric He I absorption features.
The spectrum is dominated by intense emission lines. They originate in a dense emission region, indicated by Fe II emission.}
\label{spectra_final2}
\end{figure*}
\begin{table}
\label{atomic_transitions}
\caption[]{Observed strong features in the optical spectrum of ISOSS J 20298+3559-IRS1.
The columns are: (1) transition, (2) observed peak wavelength, (3) equivalent width, (4) line flux.}
\begin{tabular}{l|c|c|c}
\hline
\hline
Line &  Obs. Wavelength & EW & Flux \\
     &    [\AA]   &   [\AA]   &       [10$^{-16}$ erg cm$^{-2}$ s$^{-1}$]\\
 \hline
H$\alpha$ & 6562.8 & -200  & 1950 \\
H$\beta$  & 4861.4 & -19.9  & 99.8\\
H$\gamma$ & 4340.4 & -5.2  & 15.2\\
HeI & 6678.15 & 0.55 &   -13.8 \\
HeI & 7065.18 & 0.29 &   -8.5  \\
FeII & 4923.06 & -2.8 & 19.2\\ 
FeII & 5017.36 & -3.5 & 19.2\\ 
FeII & 5316.05 & -4.3 & 48.3\\
FeII & 5167.75 & -3.4 & 33.4\\
FeI  & 6358.56 & -0.2 & 4.8\\
MgI(+FeI)  & 6317.84 & -1.3 & 26.8 \\ 
DIB & 6613.81 & 0.3 & -6.5\\
NI & 6442.02 & -0.4 & 9.4\\  
NI & 6491.36 & -0.8 & 16.9\\ 
NI & 6505.58 & -0.8 & 18.1\\

\noalign{\smallskip}
\hline
\noalign{\smallskip}
\noalign{\smallskip}
\end{tabular}\\
\label{atomic_transitions}
\end{table}

\subsection{The embedded Herbig B2 star IRS1}
In order to establish the pre-main-sequence nature of the bright source IRS1 we have obtained optical long-slit spectra presented in 
Fig. \ref{spectra_final2}. The spectra display strong emission lines of H\,I, Fe\,II, N\,I, Fe\,I which are all characteristic
for luminous young stellar objects (Cohen \& Kuhui 1979). Parameters of important spectral features are given in Tab. 
\ref{atomic_transitions}.

We have several strong indications that IRS1 is a rare example of an early B type pre-main-sequence object:
He I absorption lines of presumably photospheric origin have been detected at 6678 \AA \- and 7065 \AA.
Comparing their strength with the theoretical values for stellar atmospheres given by Auer \& Mihalas (1973) we derive a spectral type 
of B2 (T$_{eff}$ = 20000 $\pm$ 2500 K, log g = 4.0).
This is consistent with the presence of NI emission lines which are confined to hotter Be type stars due to their high-ionization energy 
(14.5 eV) and the large line excitation ($\sim$ 10 eV) (Andrillat et al. 1988).
The relative strength of FeII with respect to neutral FeI also requires the presence of a more ionized line-emitting region than
associated with classical T-Tauri stars (Hamann \& Persson 1992), where FeI lines often rival the strength of FeII.  
In order to derive the rotational velocity of IRS1 from the He I lines we have used the rotational broadening function given by
Gray (1976). We obtain v sin(i) = 158 $\pm$ 25 km/s, in agreement with a pre-main-sequence star of intermediate mass (Finkenzeller 1985).
A small nebulosity (size 15 $\times$ 8 arcsec$^{2}$, PA = 40$^{\circ}$) around IRS1 which is visible on POSS-B and -R plates 
was covered by our long slit spectra and identified as a reflection nebula. Its presence and the association with the dark 
cloud C1 make the emission line object IRS1 by definition a Herbig Be star (Herbig 1960). According to the most comprehensive catalogue of 
Herbig Ae/Be stars compiled by Th\'e et al. (1994), only 20 objects known to date have a spectral type B2 or earlier.

IRS1 (=MSX5C\_G075.5314-01.8159) was detected by MSX at 8.3, 12.1 and 14.7 $\mu$m and is included in the MSX 
point source catalogue (Egan et al. 1999). The source is unresolved by the small MSX telescope (beam $\sim$ 12 arcsec).
In order to explore the nature of the thermal excess emission of warm dust around IRS1 we have obtained 
diffraction-limited sub-arcsecond images with the thermal infrared camera MAX at UKIRT.
The observations cover the M, N and Q bands as well as narrow-bands centered on the 9.7 $\mu$m silicate feature and two strong
PAH bands accessible from ground at 8.7 and 11.6 $\mu$m.

\begin{figure}
\psfig{figure=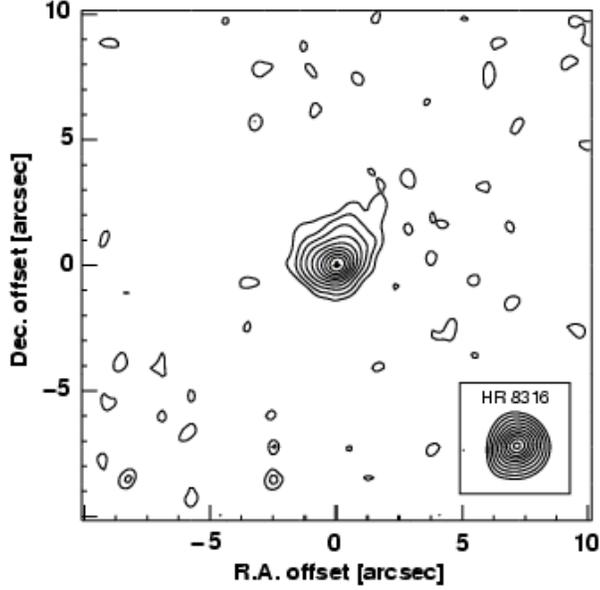,width=8.0cm,clip=true}
\caption{Diffraction-limited 11.6 $\mu$m image of the Herbig B2 star IRS1, obtained with the thermal infrared camera MAX at UKIRT
using a narrow PAH-filter. The bulk of the emission originates from an unresolved compact source (size $\le$ 0.5 arcsec FWHM).
A faint region of extended PAH emission is seen towards the north. An image of the stellar point source HR 8316 is inserted.}
\label{mir_imaging}
\end{figure}

The high-resolution data confirm the presence of a compact mid-infrared source coinciding with IRS1. An
unresolved point source (size $\la$ 0.5 arcsec FWHM) accounts for the bulk ( $\sim$ 80 \%) of emission.
The image obtained in the 11.6 $\mu$m PAH band shown in Fig. \ref{mir_imaging} shows the presence of a faint extended emission
region towards the north of the stellar source. 
Optical broadband photometry of the source in the Johnson BVRI bands have been performed in order to cover  the spectral energy distribution
of the stellar photosphere of IRS1. Tab.  \ref{photometry_herbig} presents our optical and mid-infrared photometry of IRS1.

\begin{table}
\caption[]{Optical and mid-infrared photometry of ISOSS J 20298+3549-IRS1 
given in magnitudes.}
\center{
\begin{tabular}{l|ccccccccc}
\hline
\hline
Band & B & V & R & M & P8 & S9 & N & P11 & Q\\
\hline
Flux & 17.3 & 15.7 & 14.5 & 5.7 & 3.9 & 4.1 & 3.7 & 3.1 & 2.6\\
$\sigma$ & 0.1 & 0.1 & 0.1 & 0.2 & 0.2 & 0.2 & 0.2 & 0.2 & 0.2\\
\hline
\noalign{\smallskip}
\end{tabular}}
\label{photometry_herbig}
\end{table}

\section{Discussion}

\subsection{Evidence for a pre-protostellar core in the center of the dark cloud}
Our submillimeter observations towards the central region of the dark cloud complex
have revealed the presence of a centrally condensed, dense (n(H$_{2}$) = 2 $\cdot$ 10$^{5}$ cm$^{-3}$), very cold (T$_{d} \sim$ 12 K) 
and compact (R $\sim$ 0.2 pc) core with a decreasing temperature gradient towards the center.
Similar conditions have been recently observed in FIR studies of nearby pre-protostellar cores 
(Juvela et al. 2001, Ward-Thompson et al. 2002) and are considered as the initial conditions
which pertain in clouds from which stars form. Most of the known prestellar cores are however
of low-mass. Is it possible that FIR1 contains a rare example of a massive pre-protostellar core. 
Even considering that probably more than 50 \% of the mass are dispersed during the early phases
of star formation, the total mass of the core M = 50 M$_{\odot}$ is likely to be massive enough to form
at least one late O- or early B-type star.  

By definition a pre-protostellar object has to be gravitationally bound. In order to investigate the
dynamical state of the core, we consider the virial theorem.
The condition for a gravitationally bound molecular clump or core is given by 
\begin{equation}
E_{mag} + E_{pot} = 2 \cdot (E_{kin} - E_{ext}),
\end{equation}
where E$_{mag}$ denotes the net magnetic energy, E$_{pot}$ is the gravitational energy, E$_{kin}$ the total
kinetic energy and E$_{ext}$ accounts for external pressure.
We assume that kinetic energy consists of the thermal energy of the molecules and the energy of
their turbulent motion:
\begin{equation}
E_{kin} = E_{therm} + E_{turb} = \frac{3}{2} N k T  + \frac{3}{2} M \sigma_{turb}^{2}
\end{equation}
We derive $\sigma^{2}_{turb} = \Delta V^{2} / (8 ln 2) - k T/m$, where  $\Delta V$ is the observed line width
of our ammonia data and m the mass of the ammonia molecule. The total line width of $\Delta V$ = 0.955 km s$^{-1}$ 
is an upper limit, since it was derived from a gaussian fit of the whole maingroup. 
On the basis on the (1,1) spectrum
presented in Fig. \ref{effel_nh3} it is unclear if the two peak components correspond to the resolved hyperfine structure
transition, which would in term lower $\Delta V$, or if the double peak is simply due to noise in the spectrum.
The error of the kinetic temperature $T_{kin}$ = 21.3 $\pm$ 5 K is rather high, due to the low SNR in the (2,2) spectrum.
Usually the kinetic gas temperatures in cold and dense regions are lower than the obtained dust temperatures.
We therefore consider T$_{kin}$ as an upper limit.

With the effective core radius of R = 0.2 pc we derive a gravitational potential of the homogenous ellipsoid (Liljestr\"om 1991)
\begin{equation}
E_{pot} = 3 G M^{2} / 5 R
\end{equation}

We find $E_{pot} = -6.4 \cdot 10^{37} J$,  $E_{therm} =  1.0 \cdot 10^{37} J$ and
$E_{turb} = 2.5 \cdot 10^{37}$. Relative to $|E_{pot}|$ the total kinetic energy sums up to 52 \%.
Neglecting external pressure and magnetic energy for which we have no data at present, we conclude
that the cold core is roughly in virial equilibrium. If it will collapse or not, 
either as a whole or in fragments, will depend mainly on the local turbulent velocity field. Further molecular 
line data of sufficiently high spatial resolution are needed to assess the collapse issue. 
Tab. \ref{properties_core} summarizes the properties of the cold core region.

\begin{table}
\
\caption[]{Physical properties of the cold core in ISOSS J 20298+3559.}
\center{
\begin{tabular}{|l|c|l|}
\hline
Mass  &  50 $\pm$ 10 & M$_{\odot}$\\
Dust Temperature & 12 $\pm$ 2 & K\\
Size & 0.55 $\times$ 0.29 & pc$^{2}$\\
Visual extinction A$_{V}$ & 30 & mag \\
Column density n(H$_{2}$)&  2.2 $\cdot 10^{22}$ & cm$^{-2}$ \\
Density N(H$_{2}$)& $\sim$ 2 $\cdot 10^{5}$ & cm$^{-3}$ \\
Gravitational energy & 6.4 $\cdot 10^{37}$ & J \\
\noalign{\smallskip}
\hline
\noalign{\smallskip}
\end{tabular}}
\label{properties_core}
\end{table}

\subsection{The nature of the two cold submillimeter sources SMM1 and SMM3}
Phenomenologically, the youngest protostars (Class 0 objects) are characterized by three attributes (Andr\'e et al. 1993):
\begin{enumerate}
\item Very little emission at $\lambda < 10 \mu$m,
\item Spectral energy distribution similar to a blackbody of T $\sim$ 15-30 K and
\item L$_{submm}$ / L$_{bol}$ $>$ 5 $\times$ 10$^{-3}$, where  L$_{submm}$ is the luminosity measured at  $\lambda > $ 350 $\mu$m.
\end{enumerate}
\begin{figure*}
\psfig{figure=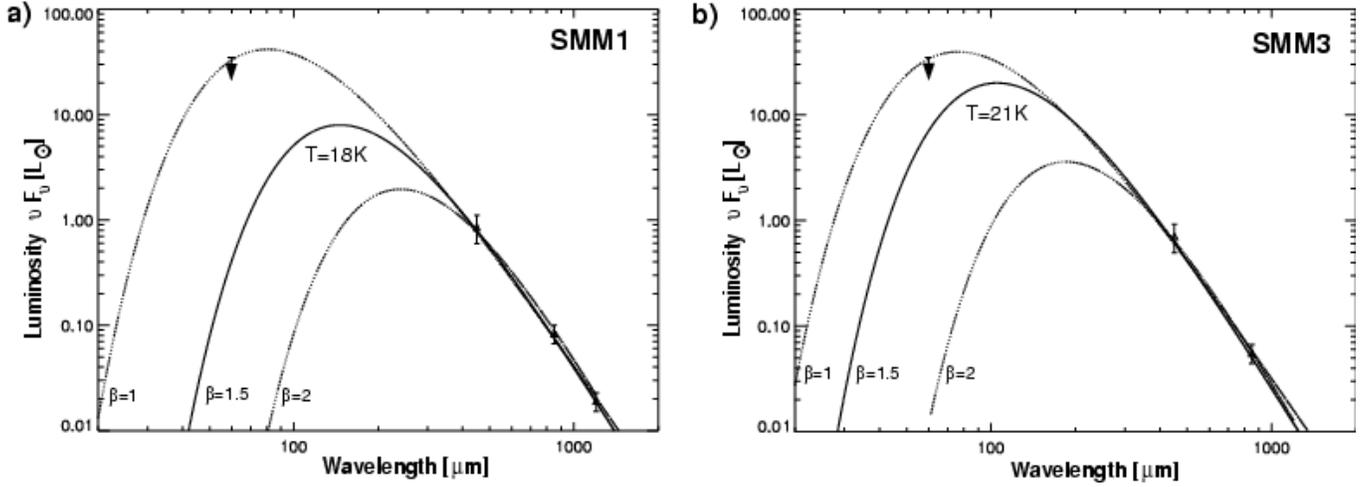,width=\linewidth,clip=true}
\caption{Spectral energy distribution of the compact submillimeter sources SMM1 and SMM3. The temperature is only indicated
for $\beta$ = 1.5. See text for a discussion of the value of the $\beta$-parameters. The arrow indicates a 60\,$\mu$m upper
limit based on IRAS photometry.}
\label{smm}
\end{figure*}
We suggest that SMM1 and SMM3 fullfil these criteria.
SMM1 and SMM3 are not detected in the mid-infrared, making them candidates
for the youngest protostellar objects in the vicinity of ISOSS J 20298+3559.

The spectral energy distributions of SMM1 and SMM3 are presented in Fig. \ref{smm}.
Relying only on our submillimeter photometry, the spectral energy distributions of both sources
can be described by modified blackbodies with dust temperatures between 10 and 38 K.
The large uncertainty of the temperature is mainly due to the unknown dust emissivity index $\beta$ which covers
the range between 1.0 $< \beta <$ 2.0 for protostellar objects and YSOs (Dent et al. 1998). 
To unambiguously determine the dust emissivity $\beta$, measurements at the maximum or Wien part of the energy distribution
are required.

Next we try to provide 60 $\mu$m upper limits for SMM1 and SMM3, considering the flux of FIR1 minus that of main contributors. 
Due to confusion with large amounts of distributed cold dust detected by the large-beam IRAS 100 $\mu$m and 
ISOPHOT 170\,$\mu$m bands, the photometry at these wavelengths can hardly be used for a reliable characterization of
the two faint compact sources. The most stringent upper limit for SMM1 and SMM3 can be derived from the IRAS 60 $\mu$m band.
The photometry at this wavelength is however affected by the emission of very small dust particles, transiently
heated in the reflection nebula of the Herbig B2 star IRS1.
The presence of such extended PAH and VSG emission is indicated on our mid-infrared images of IRS1. The IRAS measurements at 
12$\mu$m, 25$\mu$m and 60 $\mu$m are particularly sensitive to such an extended emission component. 
In order to determine the contribution of thermal radiation of transiently heated PAH and very small grains in these bands
we have used the three-phase dust models by D\'esert et al. (1990).
After subtraction of compact emission from IRS1 according to our model (see next section), we determined
the relative contributions of PAH and VSG emission from the residual IRAS 12$\mu$m and 25$\mu$m photometry and using
the model  by D\'esert et al. (1990) for the radiation field of a stellar source with T$_{eff}$ = 20 000K.
We find a contribution of F$_{60\mu m}$ = 7 $\pm$ 3 Jy from the Herbig star IRS1 and its reflection to the total flux density of F$_{60\mu m}$ = 14 Jy.
The residual flux density of F$_{60\mu m}$ = 7 Jy is considered as an upper limit for each of the two sources.

Using the possible range 1.0 $< \beta <$ 2.0 for the emissivity index, we derive total masses of 3.7 M$_{\odot}$ $<$ M $<$ 18 M$_{\odot}$ for SMM1 and 2 M$_{\odot}$ $<$ M $<$ 8 M$_{\odot}$ for SMM3.
The corresponding bolometric luminosities cover the range 36 L$_{\odot}$ $>$ L$_{bol}$ $>$ 1.8 L$_{\odot}$ for SMM1 and 34 L$_{\odot}$ $>$ L$_{bol}$ $>$ 3.3 L$_{\odot}$ 
for SMM3 respectively.
If we favor an intermediate emissivity index of $\beta$ $\sim$ 1.5, which was the average value found in the large submillimeter survey of 73 young stellar objects by Dent et al. (1998),
then we obtain for the two sources the following masses and luminosities:\\
SMM1: M = 8.0 M$_{\odot}$, L$_{bol}$ = 7 L$_{\odot}$, T$_{d}$ = 18 K\\
SMM3: M = 3.3 M$_{\odot}$, L$_{bol}$ = 17 L$_{\odot}$, T$_{d}$ = 21 K\\

We find L$_{submm}$ / L$_{bol}$ $\ga$ 1 \% for each choice of 1.0 $< \beta <$ 2.0,
making  SMM1 and SMM3 bona-fide candidates for Class 0 objects.
Further evidence for SMM1 and SMM3 being genuine protostars could be provided via detections of a molecular outflow 
and/or a cm-wavelength continuum source.

\subsection{Evidence for accretion onto the extremely young Herbig B2 star IRS1}

\begin{figure}[]
\psfig{figure=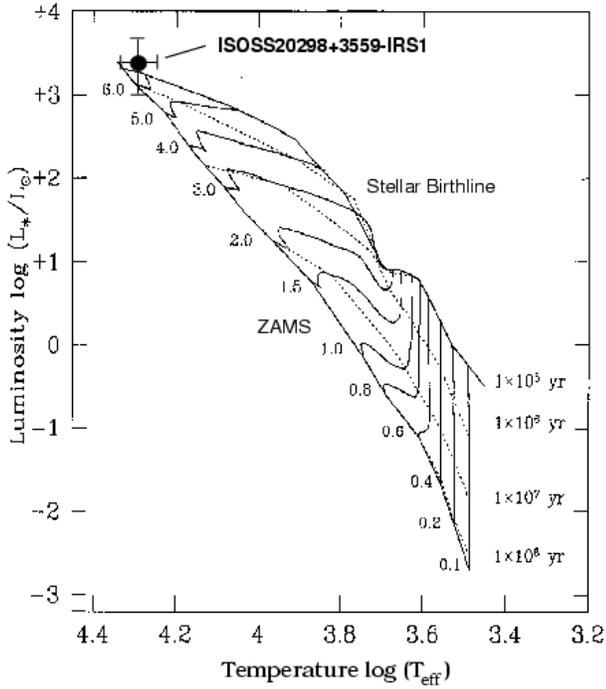,width=\linewidth,clip=true}
\caption{
H-R diagram with stellar effective temperatures and luminosities, adopted from Palla \& Stahler (1999).
Stellar masses along the zero-age main sequence (ZAMS) are indicated in units of M$_{\odot}$.
The Herbig B2 star IRS1 is located close to the stellar birthline, which was calculated for an accretion
rate of  $\dot{M}$ = 10$^{-5}$ M$_{\odot}$ yr$^{-1}$. The pre-main-sequence tracks overlaid
indicate a stellar age of less than 40000 yr.}
\label{palla}
\end{figure}

The Herbig B2e star IRS1 is the most luminous young stellar object embedded in the cold cloud complex.
A very early evolutionary stage of IRS1 is indicated from its locus in the Hertzsprung-Russel diagram presented in Fig.
\ref{palla}. According to the pre-main-sequence evolutionary tracks by Palla \& Stahler (1993) the source is located close
to the stellar birthline computed for $\dot{M}$ = 10$^{-5}$ M$_{\odot}$ yr$^{-1}$ corresponding to a M$_{star}$ =  6.5 M$_{\odot}$
star with an age of less than 40000 yr immediately after becoming optically visible.

\begin{figure}
\center{\psfig{figure=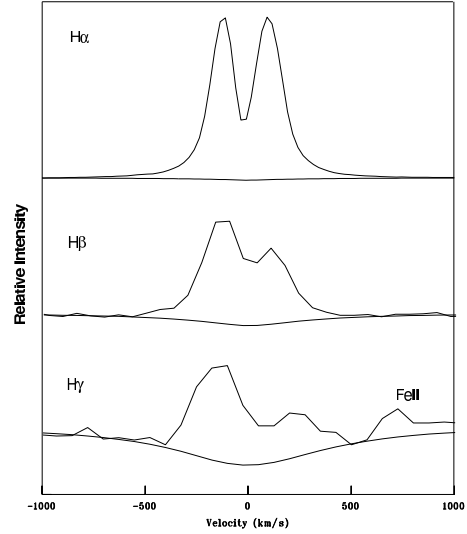,angle=0,width=6cm,clip=true}}
\caption{Emission line profiles of the H I Balmer series.
H$\alpha$,  H$\beta$ and H$\gamma$  are shown from top to bottom.
The stellar continuum for a standard star of identical spectral
type is shown for each transition.
In addition to a strong central self absorption,
the redshifted emission line component is suppressed for the higher
order lines, indicating infall in an accretion process.}
\label{balmer}
\end{figure}

The observation of inverse P Cygni profiles as shown by assymetric red and blue
line components of the H I Balmer transitions presented in  Fig. \ref{balmer} provides
evidence for ongoing mass infall in a magnetospheric accretion process (Edwards et. al 1994).
From the redshifted absorption component in the Si\,II lines at 6347.1 \AA \- and 6371.4 \AA 
\- transitions shown in Fig. \ref{spectra_final2} we derive an infall speed of 110 $\pm$ 10 km s$^{-1}$. 

Near-infrared photometry and our sub-arcsecond resolution mid-infrared images revealed the 
presence of strongly condensed (r $\la$ 250 AU) warm dust in the immediate surrounding of IRS1. 
In order to model the optical and infrared spectral energy distribution of the heavily reddened source IRS1,
a reliable determination of the extinction towards the source is essential.
Based on the effective stellar temperature derived from our observations of photospheric He I lines,
we apply a Kurucz model atmosphere with T$_{eff}$ = 20 000 K for the stellar photosphere. We assume solar
metallicity. Our dereddened BVRI photometry is best described for A$_{V}$ = 6.1 $\pm$ 0.3 mag where the error is due to the
uncertainty of T$_{eff}$. Assuming the distance d = 1.8 kpc of the cloud complex we derive a stellar radius
of R$_{\star}$ = 4 $\pm$ 0.5 R$_{\odot}$ from the surface flux density of the Kurucz model. The integrated bolometric
luminosity of the photosphere is L$_{\star}$ = 2200 $\pm$ 600 L$_{\odot}$.

\begin{figure}
\psfig{figure=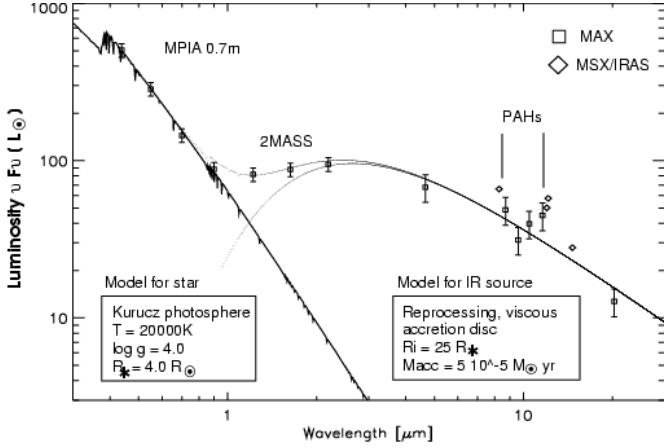,width=\linewidth,clip=true}
\caption{Dereddened optical and infrared spectral energy distribution of the Herbig star IRS1.
We have modelled the strong infrared excess emission by an optically thick accretion disk.
The star is described by a Kurucz model with a T$_{eff}$=20 000K photosphere (log g=4.0). Our observations
obtained with the MPIA 0.7m telescope in the optical and with MAX at UKIRT in the thermal infrared
are in good agreement with the model.}
\label{herbig_accretion1}
\end{figure}

How can our observations of ongoing accretion and circumstellar dust for this extremely young stellar object
be combined in a consistent model?
Fig. \ref{herbig_accretion1} shows the dereddened spectral energy distribution of IRS1.
The near-infrared emission bump seen at about 3 $\mu$m wavelength is characteristic for Herbig AeBe stars and has been interpreted
either by the presence of a circumstellar disk with an inner hole (Hillenbrand et al. 1992), by spherical envelopes
(Hartmann et al. 1993) or the combination of both (Miroshnichenko et al. 1999).
While there is general consensus that the Herbig Ae stars are surrounded by circumstellar disks similar to
those of T Tauri stars during most of their pre-main-sequence phase (Natta et al. 2001), only one out of seven Herbig Be
stars was detected by mm-interferometry (Natta et al. 2000). Extended halos and envelopes in the 
near- (Leinert et al. 2001) and mid-infrared (Prusti et al. 1994) are more frequently observed around
more massive Herbig Be stars. However, these observational differences could be due to a faster evolution of 
massive stars and differences in the relative timescales of the PMS phases, which limit
the detection of disks around early B stars to the very early stage of their evolution.

We suggest that our spectroscopic signatures of ongoing accretion and the infrared excess can best be
combined in terms of an accretion disk around IRS1.
We have calculated the spectral energy distribution of an
active viscous accretion disk including radiative heating by photospheric radiation from IRS1.
The radial temperature dependence of the optically thick, geometrically thin disk is then assumed to 
be $T(r) = [T_{acc}(r)^{4} + T_{rad}(r)^{4}]^{1/4}$.
We use a standard $\alpha$-disk (Lynden-Bell \& Pringle 1974). Its accretion disk temperature profile is given by
\begin{equation}
T_{acc}(r) = \left [ \frac{3 G M_{\star} \dot{M}}{8 \pi \sigma r^{3}} \left ( 1 - \sqrt{\frac{R_{\star}}{r}} \right ) \right ]^{1/4}.
\end{equation}
In the flat disk approximation its temperature profile due to irradiation only is
\begin{equation}
T_{rad}(r) = \left [ \frac{2 T_{eff}^4 R_{\star}^{3}}{3 \pi r^{3}} \right ]^{1/4}.
\end{equation}

Emerging spectral energy distributions have been computed for varying inner disk radii R$_{i}$ and accretion rates $\dot{M}_{acc}$.
As shown in Fig. \ref{herbig_accretion1} the fit of the star+disk SED to the dereddened photometry of IRS1 is consistent
with a dusty, optically thick accretion disk. Tab. \ref{properties_mir1} summarizes the model parameters of the system.

\begin{table}
\caption[]{Properties of the Herbig B2 star IRS1 and its accretion disk.}
\center{
\begin{tabular}{|l|c|l|}
\hline
Mass  &  6.5 $\pm$ 1 & M$_{\odot}$\\
Luminosity & 2200 $\pm$ 600 & L$_{\odot}$\\
Effective temperature & 20000 $\pm$ 2500  & K\\
Radius & 4.0  $\pm$ 0.5 & R$_{\odot}$ \\
Visual extinction A$_{V}$   & 6.1  $\pm$ 0.3 & mag \\
Age &    $<$ 40000 & yr \\
Rotational velocity & 158  $\pm$ 25 & km/s \\
Inner disk radius  & 60.0 & R$_{\odot}$\\
Accretion rate & 5 $\times$ 10$^{-5}$ & M$_{\odot}$ yr$^{-1}$\\
\noalign{\smallskip}
\hline
\noalign{\smallskip}
\end{tabular}}
\label{properties_mir1}
\end{table}

Next we consider whether our data could also allow an envelope:
Our spectrophotometry of IRS1 obtained in the narrow-band silicate filter at 9.7 $\mu$m shows a weak absorption feature 
($\tau_{9.7 \mu m} \sim$ 0.2 $\pm$ 0.1) which is consistent with the total line-of-sight extinction towards the source
according to Whittet el al. (1987).
Models of spherical envelopes which account for the observed spectral energy distributions of Herbig Be stars 
all predict strong silicate emission (Berilli et al. 1992, Miroshnichenko et al. 1997). 
In order to explain the suppression of the silicate emission feature in a B0 star, the inner radius of the envelope must
be about 0.02 pc (Natta et al. 2000). This is inconsistent with our diffraction-limited mid-infrared images, which show 
that the bulk of flux from IRS1 arises from a true point source, i.e. the radius of the mid-infrared emitting region is $\rm <$ 400 AU.

\subsection{Star formation efficiency}

We estimate the star formation efficiency SFE = M$_{stars}$ / M$_{tot}$ in the central region ISOSS J 20298+3559-FIR1.
M$_{tot}$ includes the masses of the diffuse component, the dense core and all young stellar objects.
We have determined the masses for the near-infrared objects IRS2..5 from the absolute magnitude-mass relation 
by Carpenter et al. (1993).

Concerning the masses of the Class 0 sources SMM1 and SMM3 we assume that they
loose 50 \% of their envelopes during their further evolution (Bachiller et. al. 1996).
The mass of all young stellar objects is then M$_{stars}$= 20 M$_{\odot}$. Adding the mass of gas and dust in FIR1 to the stellar
masses we obtain M$_{tot}$ = 140  M$_{\odot}$. This results in a star formation efficiency of about 14 \%.

To compare: The SFE is about 2-9 \% for the Taurus complex (Cohen \& Kuhi 1979) and 1-9 \% in Orion A+B (Carpenter 2000).
ISOSS J 20298+3559-FIR1 clearly has at least the efficiency of these low- and high-mass star forming regions.

\section{Conclusions}
We have identified the young star forming region ISOSS J 20298+3559
performing a cross-correlation of cold compact far-infrared sources from the ISOPHOT
170 $\mu$m Serendipity Survey database with the 2MASS, MSX and IRAS surveys.
Multi-wavelength follow-up observations of this region yield:

\begin{enumerate}
\item The star forming region is associated with a complex of four optical dark clouds C1..C4
which have a total mass of 760 M$_{\odot}$.

\item We derived a distance of 1800 $\pm$ 300 pc based
on optical extinction data. This associates the region
with the Cygnus\,X Giant Molecular Cloud in agreement
with our molecular line kinematics.

\item The cold ISOSS source FIR1 corresponds to the dense inner
region of the central dark cloud C1 and contains a total mass of 120 M$_{\odot}$ gas and
dust with an average temperature of 16 K.

\item We have identified two candidate Class 0 objects SMM1 and SMM3. The
sources have masses of 8 and 3.5  M$_{\odot}$ which makes them precursors of intermediate mass stars.

\item The externally heated cloud core of C1 has a total mass of 50 M$_{\odot}$
and a central dust temperature as low as 11K. Ammonia in the NH$_{3}$(1-1) transition has
been detected. The object is gravitationally bound as derived from our ammonia molecular line observations,
which makes it a candidate massive pre-protostellar core.

\item The most luminous object in the vicinity is the Herbig B2 star IRS1. 
The object has a mass of 6.5 M$_{\odot}$ and a bolometric luminosity of
2200 L$_{\odot}$. Inverse P Cygni profiles in the higher HI Balmer series 
and SiII lines indicate ongoing accretion. The spectral energy is well described by a Kurucz model for the
stellar photosphere and a viscous reprocessing accretion disk. The stellar age inferred from
pre-main-sequence evolutionary tracks is less than 40000 yr.

\item Several embedded near-infrared sources have been identified. One of them (IRS7) is surrounded
by a mid-infrared reflection nebula.

\item The star formation efficiency in the dense and cold region FIR1 is about 14 \%.

\end{enumerate} 

\acknowledgements
We thank the referee, Debra Shepherd, for useful comments.
The ISOPHOT Data Center at MPIA is supported by Deutsches Zentrum f\"ur Luft- und
Raumfahrt e.V. (DLR) with funds of Bundesministerium f\"ur Bildung und Forschung,
grant No. 50QI0201. OK thanks the Wernher von Braun-Stiftung zur F\"orderung der
Weltraumwissenschaften e.V. for financial support.
This study made use of the SIMBAD database operated at CDS, Strasbourg, France.
HIRES images were provided by the Infrared Processing and Analysis Center.
This publication makes use of data products from the Two Micron All Sky Survey, 
which is a joint project of the University of Massachusetts and the Infrared Processing 
and Analysis Center/California Institute of Technology, funded by the National Aeronautics 
and Space Administration and the National Science Foundation.
This research made use of data products from the Midcourse Space 
Experiment, funded by the Ballistic 
Missile Defense Organization with additional support from NASA 
Office of Space Science.

\end{document}